\documentclass[aip,apl,reprint,showpacs]{revtex4-1}
\usepackage{verbatim}
\usepackage{textcomp}
\usepackage{graphicx}

\begin{document}


\title{Detection of variable tunneling rates in silicon quantum dots} 


\author{A. Rossi}\email[Electronic mail: ]{ar446@cam.ac.uk}
\author{T. Ferrus}\affiliation{Hitachi Cambridge Laboratory, J.J. Thomson Avenue, Cambridge, CB3 0HE, U.K.}


\author{W. Lin}\affiliation{Tokyo Institute of Technology, Quantum Nanoelectr. Res. Ctr., Meguro Ku, Tokyo 1528552, Japan}
\author{T. Kodera}
\affiliation{Tokyo Institute of Technology, Quantum Nanoelectr. Res. Ctr., Meguro Ku, Tokyo 1528552, Japan}
\affiliation{Institute for Nano Quantum Information Electronics, the University of Tokyo, 4-6-1, Komaba, Meguro, Tokyo, Japan}
\affiliation{PRESTO, Japan Science and Technology Agency (JST), Kawaguchi, Saitama 332-0012, Japan}
\author{D.A. Williams}
\affiliation{Hitachi Cambridge Laboratory, J.J. Thomson Avenue, Cambridge, CB3 0HE, U.K.}
\author{S. Oda}
\affiliation{Tokyo Institute of Technology, Quantum Nanoelectr. Res. Ctr., Meguro Ku, Tokyo 1528552, Japan}



\date{\today}

\begin{abstract}
Reliable detection of single electron tunneling in quantum dots (QD) is paramount to use this category of device for quantum information processing. Here, we report charge sensing in a degenerately phosphorus-doped silicon QD by means of a capacitively coupled single-electron tunneling device made of the same material. Besides accurate counting of tunneling events in the QD, we demonstrate that this architecture can be operated to reveal asymmetries in the transport characteristic of the QD. Indeed, the observation of gate voltage shifts in the detector's response as the QD bias is changed is an indication of variable tunneling rates.
\end{abstract}


\maketitle 

\begin{figure}[]
\includegraphics[scale=0.5]{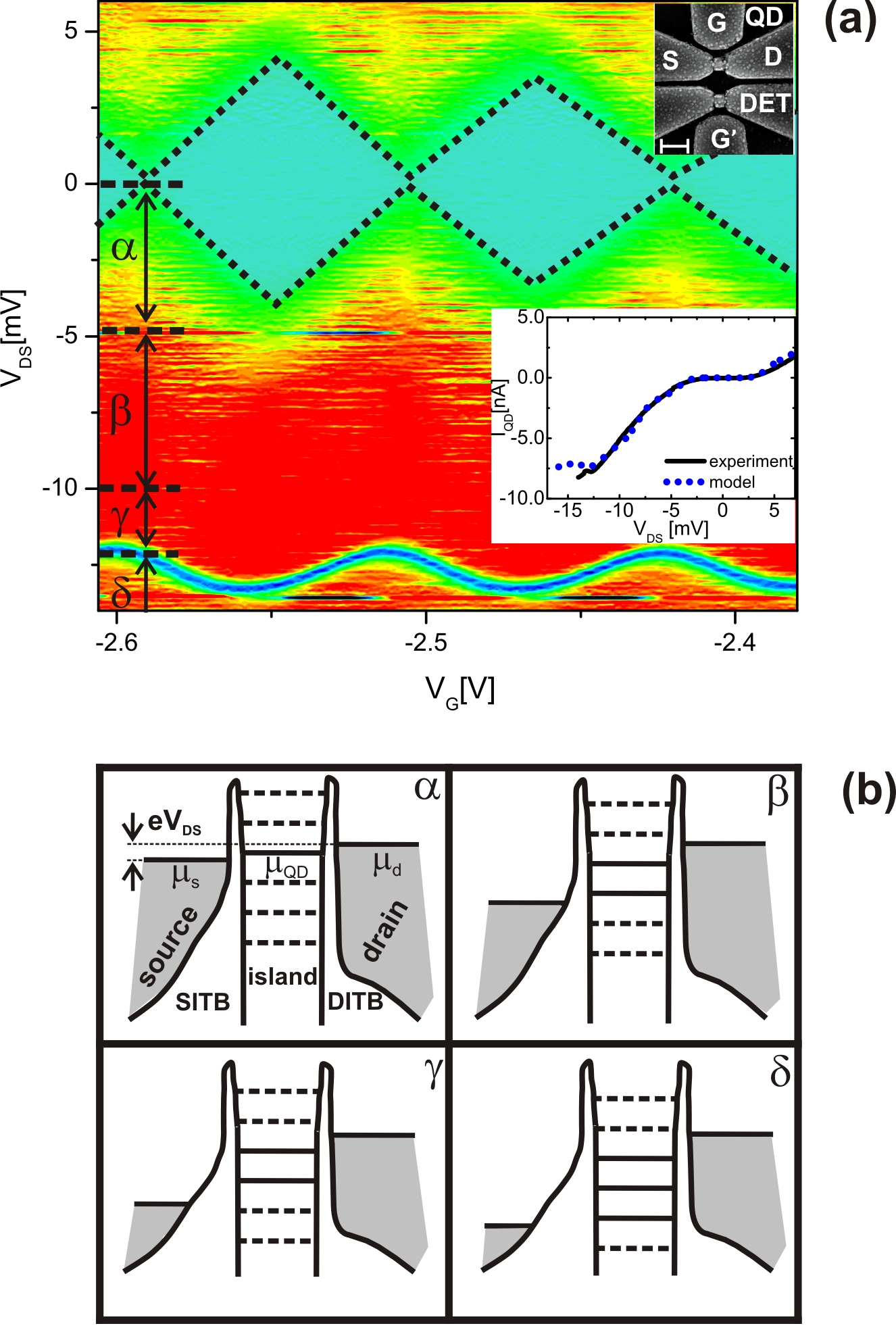}
 \caption{\label{fig:NDC}a) QD stability plot. Differential conductance varies from -0.5 $\mu$S to +1.0 $\mu$S in the scale range reported. Negative values of the differential conductance define the NDC feature at negative V$_{DS}$. Top inset: SEM image of a representative device showing both QD and detector. Scale bar is 200 nm. Bottom inset: experimental (solid line) and calculated (dotted line) IV characteristic of the QD at V$_G$=-2.55V. b) Schematic energy level diagrams showing the origin of the NDC in the QD for increasingly negative  V$_{DS}$. The lettering indicates which region of the stability plot in (a) each diagram refers to. Solid (dashed) lines in the island represent levels available (unavailable) for transport. Transport channels are evenly separated by the charging energy of the QD.}
 \end{figure}

\begin{figure}[t]
\includegraphics[scale=0.5]{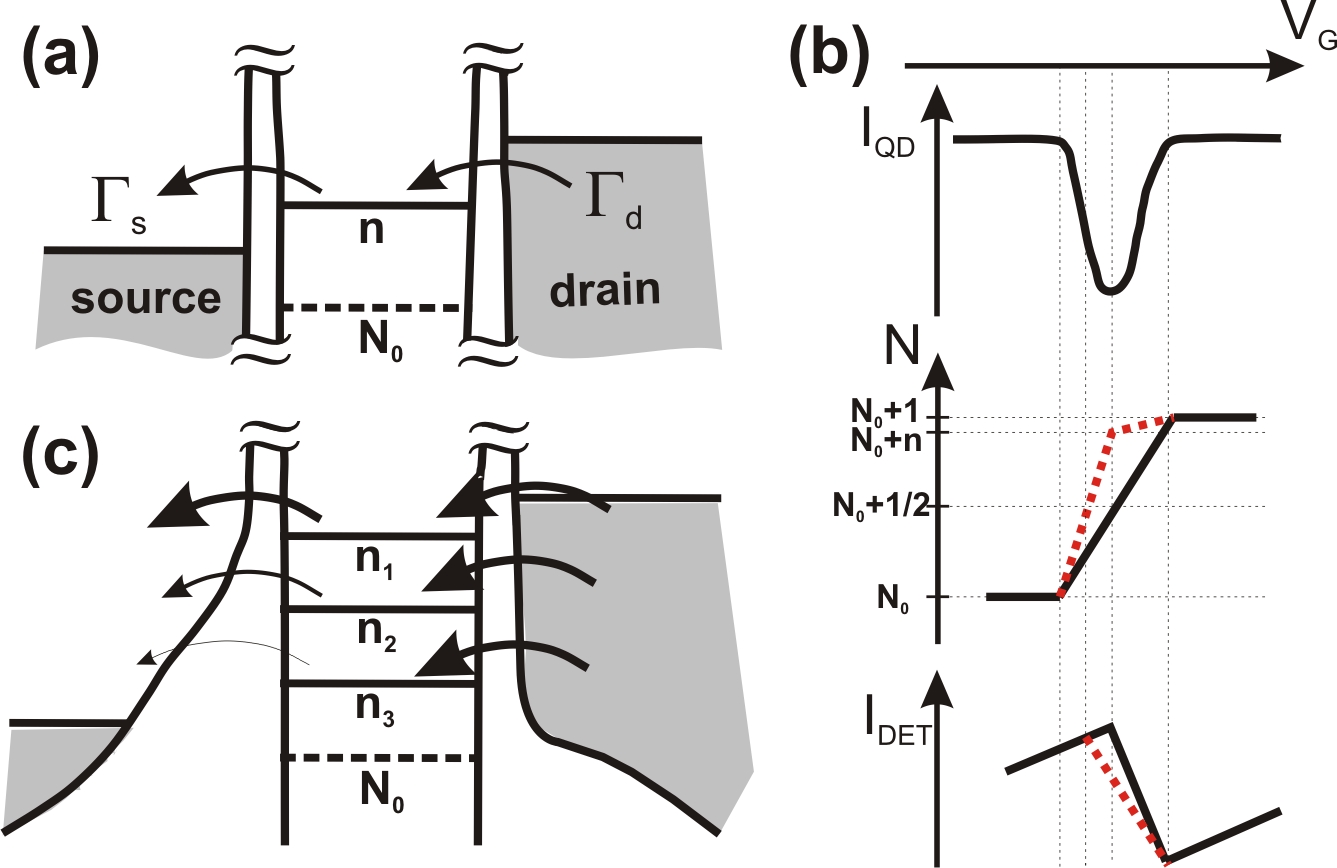}
 \caption{\label{fig:rates} a) QD energy level diagram for the bias region $\alpha$ where $\Gamma_d\approx\Gamma_s$ at the center of the transport window. b) the effect of sweeping V$_G$ (i.e. the Fermi energy of the QD) over a CB oscillation on both the average number of electrons in the QD and the current in the detector. Solid lines represent the situation  $\Gamma_d\approx\Gamma_s$ (i.e. n$\approx\frac{1}{2}$), dotted lines are for $\Gamma_d\gg\Gamma_s$ (i.e. n$\lesssim$1). c) QD energy level diagram for the bias region $\delta$. Different tunneling rate amplitudes are schematically indicated by the arrow thicknesses.}
 \end{figure}

Silicon-based quantum dot (QD) architectures~\cite{gorman,gareth,angus} for quantum computing have recently emerged as an attractive possibility in view of their scalability, their compatibility with the widely accessible complementary metal-oxide-semiconductor technology~\cite{fujiwara} and their long coherence time for electron spin states.~\cite{fujisawa} In these systems, readout of qubit states typically requires charge sensing and spin-to-charge conversion.~\cite{Hanson} Nevertheless, charge sensing is useful to achieve the few-electron regime in the dot which is instrumental to perform spin manipulations.~\cite{koppens} Demonstration of reliable single-electron detection is, therefore, crucial for the implementation of quantum logical operations and future development of complex quantum computational schemes.\\\indent
Here, we report charge sensing using two capacitively coupled single-electron tunneling devices (SET) fabricated in degenerately phosphorus-doped silicon. One SET is used as a QD weakly coupled to two electron reservoirs (namely, source and drain) while the remaining SET is used as a detector of the QD charge state. We show that modifications of the drain-source bias voltage of the QD can lead to changes in the transparency of its tunnel barrier(s). This results in the observation of a negative differential conductance (NDC) in the QD's stability plot as well as in a shift with respect to gate voltage of the expected detector's response.~\cite{nordberg} These effects may arise from some intrinsic characteristics of the material, such as defects at the Si/SiO$_2$ interface~\cite{sun} and disordered distribution of dopants in the nano-structure.~\cite{myJAP} Ultimately, they may limit the ability of detecting electrons and need to be taken into account for the reliable design of these devices.\\\indent
The top inset of Fig.~\ref{fig:NDC}(a) shows a scanning electron micro-graph (SEM) image of a device similar to those investigated.
The system material is silicon-on-insulator with a concentration of phosphorus atoms in the active layer of about $10^{19}$ cm$^{-3}$. The two SETs are trench-isolated via reactive ion etching so that no current can flow between them and coupling is only capacitive. The islands of the two SETs are electron-beam-defined by patterning constrictions in the nano-structure that act as tunnel barriers due to severe carrier depletion caused by both trapping at the Si/SiO$_2$ interface~\cite{myJAP} and dopant deactivation.~\cite{bjork} It has been previously reported that specific fabrication arrangements~\cite{mizuta} should be put in place to mitigate the effect of disorder on these devices' characteristics. Our devices were rather realised to optimize coupling effects, hence the observation of asymmetries in the transport characteristics is not surprising, as we shall discuss next. 
\\\indent In order to significantly suppress thermal fluctuations and allow Coulomb Blockade (CB) to arise, experiments have been carried out in a He$^3$ cryostat at the base temperature of 300 mK. The top SET is typically used as a QD whose electron number is modified by sweeping the voltage of the adjacent gate electrode (V$_G$). The average current flowing from source to drain through the QD is controlled by acting on the bias voltage (V$_{DS}$). The bottom SET is used as a charge detector whose operating point is selected to maximize charge sensitivity. Since the voltages applied to the QD gate to modify its electron number slightly affect the potential of the detector, simultaneous sweep of the gate electrode close to the detector (G\textquotesingle) is performed to compensate for unwanted shifts away from optimal bias. The suitable compensation ratio between the two voltages is selected according to capacitive considerations, as detailed elsewhere.~\cite{myAPL}\\\indent
The stability plot for the QD device when the detector's electrodes are grounded is shown in Fig.~\ref{fig:NDC}(a). Besides the expected diamond-shaped characteristics due to CB, the presence of a NDC for negative values of V$_{DS}$ is clearly seen. This feature's onset voltage is modulated by V$_G$ in a similar manner to the CB diamonds edges (dotted line in the figure). This implies that the electrostatic potential of the dot plays a significant role in the mechanism originating the NDC. Furthermore, we do not observe any NDC for V$_{DS}>0$ which indicates an asymmetrical transport mechanism consistent with unequal tunneling probabilities through the two tunnel barriers. The formation of these barriers is affected by the energy spread of interface traps~\cite{sze} as well as by the random spatial distribution of donors in the constricted regions.~\cite{evans} These phenomena are likely to give rise to asymmetrical bias-dependent tunneling probabilities in our device. Nonetheless, variable tunneling rates in Si nano-wires have been previously reported.~\cite{saitoh,koba} In Fig.~\ref{fig:NDC}(b) a possible mechanism to explain the origin of the NDC is sketched. Firstly, we assume a strong asymmetry in the shape of the tunnel barriers. In particular, the source-island tunnel barrier (SITB) should show energy-dependent tunneling probability whilst the drain-island tunnel barrier (DITB) should be nearly energy-independent in the bias range of interest. Secondly, it is reasonable to assume that the capacitive coupling between the QD and the source-drain electrodes affects the dot's energy levels whenever the Fermi level in either reservoir is modified via V$_{DS}$.~\cite{koba} In region $\alpha$ of the stability plot, CB is lifted because the Fermi energy in the QD, $\mu_{QD}$,(solid line in Fig.~\ref{fig:NDC}(b)) falls within the transport window between the Fermi energies of source ($\mu_{s}$) and drain ($\mu_{d}$). For more negative V$_{DS}$ (region $\beta$) the current increases monotonically because an extra level enters the transport window. By contrast, in region $\gamma$, an increasingly negative V$_{DS}$ produces a decrease of current (i.e. a NDC) because the tunneling probability is reduced given the concurrent effects of the bias-dependent SITB and the QD's potential shift (note that we assume the drain electrode to be grounded, hence a negative V$_{DS}$ would pull down both $\mu_{s}$ and $\mu_{QD}$ leaving $\mu_{d}$ unchanged). Finally, by increasing further $|V_{DS}|$ a third transport channel becomes available (region $\delta$); this compensates the increased opacity of the SITB and the current recovers the increasing trend. The bottom inset of Fig.~\ref{fig:NDC}(a) compares one of the experimental I-V$_{DS}$ curves with the calculated characteristics for a parabolic shape of SITB~\cite{saitoh} and an abrupt shape of DITB. An analytical approach~\cite{grabert} is used to calculate the current through a double tunnel junction system as a function of bias voltage. The calculation shows that a NDC arises due to the mechanism detailed above. Importantly, the value of V$_{DS}$ at the onset of the NDC depends on the parabolic coefficients of the SITB as well as on the degree of coupling between the reservoirs and the QD. In the calculation these parameters have been chosen to fit the experimental results. Since conduction by thermally activated carrier is not taken into account in this model, the agreement between experiments and calculation is expected to become less good as the bias voltage increases. This may explain the discrepancy in the region beyond the NDC.\\\indent 
\begin{figure}[]
\includegraphics[scale=0.50]{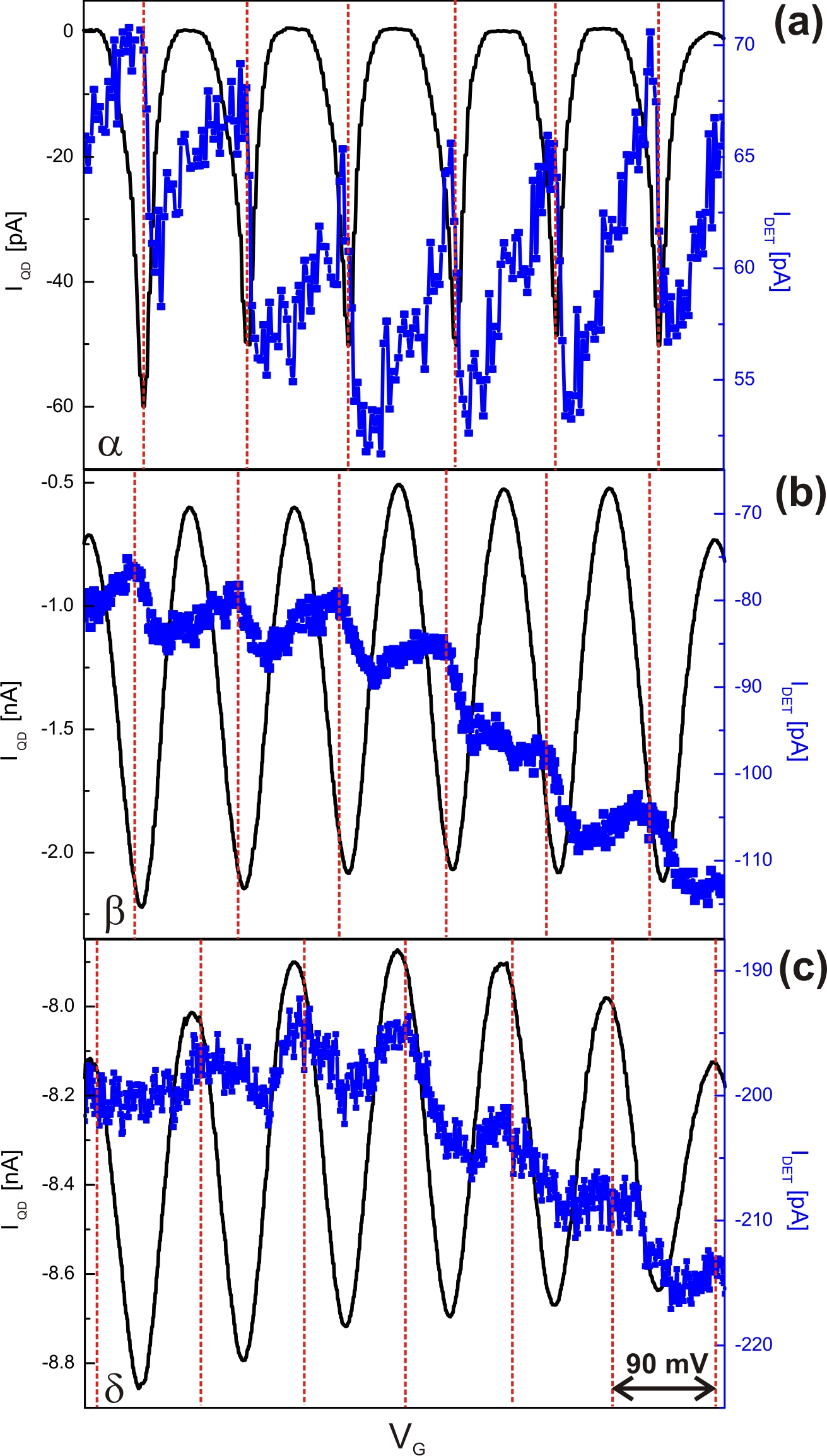}
\caption{\label{fig:shift}Simultaneous current response of the QD (smooth lines) and the detector (squared lines)  for different V$_{DS}$ as the QD's gate is swept. a)  V$_{DS}$=-2mV (region $\alpha$ of the stability diagram). Sensing steps occur at the same voltages as the CB dips. b)  V$_{DS}$=-6mV (region $\beta$ of the stability diagram). Sensing steps are shifted with respect to the CB dips. c)  V$_{DS}$=-13mV (region $\delta$ of the stability diagram). Increased shift is noticeable. The detector's bias point is chosen to maximize charge sensitivity but is not the same for all the traces.}
\end{figure}
We now turn to discuss the effect of this tunneling probability modulation on the charge sensing mechanism. In Fig.~\ref{fig:shift}, the QD and detector currents (solid black and squared blue lines, respectively) are simultaneously monitored as V$_G$ is swept for three values of  V$_{DS}$. In (a),  V$_{DS}$=-2mV which falls within region $\alpha$ of the stability plot. It can be noticed how sharp changes on the detector's trace correlate with the dips observed in the QD's current, as the dashed lines indicate. This is the clear signature of the abrupt change in the detector's electrostatic potential caused by tunnelling events of single electrons in the QD. In (b), V$_{DS}$=-6mV and we are addressing region $\beta$ of the stability plot. It is of note that now the sensing is slightly misaligned with respect to the gate voltages at which the current dips. In (c), V$_{DS}$=-13mV which is in region $\delta$, i.e. beyond the NDC. In this case, an even larger sensing shift in gate voltage can be observed. At this bias, the sensing takes place in closer proximity to a current peak than a dip. It is also worth noting that the amplitude of the CB oscillations is significantly reduced with respect to region $\beta$, despite the increase of $|V_{DS}|$. In order to improve the signal-to-noise ratio, the detector's bias point was not the same for the three graphs reported; in particular, it was occasionally needed to change the detector's drain-source voltage and this explains why I$_{DET}$ does not have always the same sign. However, this kind of modification in the detector's bias point does not affect its charge sensitivity; therefore we can confidently attribute the observed shifts to variable tunneling rate in the QD, as we shall demonstrate next.\\\indent
As mentioned earlier, in region $\alpha$ only one transport channel at a time falls within the transport window and electrons enter the QD with a rate $\Gamma_d$ and leave with a rate $\Gamma_s$ (see Fig.~\ref{fig:rates}(a)). The symmetry of the stability plot for -5mV$<$V$_{DS}<$+5mV allows us to consider  $\Gamma_d \approx\Gamma_s$   for this bias range. Under this condition, the QD's current dips whenever $\mu_{QD}$ is equidistant from the two reservoirs' Fermi levels. The average electron number, $N$, in the QD is then given by~\cite{fujiAPL}
\begin{equation}
\label{eq:N}
N=N_0+n=N_0+ \frac{\Gamma_d} {\Gamma_d +\Gamma_s}\approx N_0+\frac{1}{2}
\end{equation}
where $N_0$ is the excess number of electrons already in the dot and $n$ is the fractional number of an electron added to the QD as a result of the opening of the transport channel(s) between the drain and source electrodes. When the two barriers are almost symmetrical (e.g. in region $\alpha$), $n$ is very nearly equal to $\frac{1}{2}$ and the transition from $N_0$ to $N_0+1$ electrons in the dot is sketched in Fig.~\ref{fig:rates}(b) by the solid line. From the measurements of Fig.~\ref{fig:shift}(a) we can conclude that, since the dips of I$_{QD}$ line up with the onset of the sensing steps, these latter occur whenever the electron number in the QD is $N=N_0+\frac{1}{2}$. However, when the barriers are strongly asymmetrical $N$ approaches an integer (see eq.~(\ref{eq:N})), precisely $N_0+1~(N_0)$ for  $\Gamma_d\gg\Gamma_s~(\Gamma_d\ll\Gamma_s)$; as a consequence, the sensing points will be shifted towards smaller (larger) gate voltages. In other words, the detector's response would shift in gate voltage on the conditions   $\mu_d=\mu_{QD}$ for $\Gamma_d\gg\Gamma_s$ and $\mu_s=\mu_{QD}$ for  $\Gamma_d\ll\Gamma_s$. Fig.~\ref{fig:rates}(b) shows with red dotted lines the situation relevant to our experimental condition. Indeed, the presence of a NDC at V$_{DS}<0$ is consistent with a strong reduction of $\Gamma_s$ for increasingly negative bias; this leads to an increase of the fractional number of electrons in the QD and the consequent shift of the detector's response toward smaller V$_G$. Further confirmation of a strong decrease of tunneling probability comes from the decreasing amplitude of the CB oscillations for increasingly negative bias voltages (compare traces in Fig.~\ref{fig:shift}(b) and (c)).\\\indent
For the sake of completeness, we wish to point out that for increasingly negative V$_{DS}$, apart from making the tunnel barriers asymmetrical, more transport channels become available and the fractional number of electrons in the QD will be determined by the tunnelling rates relevant to each of these levels. Fig.~\ref{fig:rates}(c) represents the condition pertinent to region $\delta$ of the stability plot where the deeper energy levels will contribute to the higher fractional electron numbers. Indeed, by assuming again a parabolic SITB and a sharp DITB, we have
\begin{equation}
\Gamma_{1d}\approx\Gamma_{2d}\approx\Gamma_{3d}\approx\Gamma_{1s}\gg\Gamma_{2s}\gg\Gamma_{3s}
\end{equation}
$\Gamma_{id(s)}$  being the tunneling rate of the $i^{th}$ channel to drain (source).
This would lead to
\begin{equation}
n_1\approx\frac{1}{2}<n_2<n_3\lesssim 1
\end{equation}

$n_i$ being the fractional electron number of the $i^{th}$ channel. Clearly, the overall effect in terms of detector's response shift would be comparable to the one described for the single transport channel.
\\\indent In conclusion, we have reported charge sensing of a QD by a capacitively coupled single-electron transistor. The QD has revealed an asymmetrical characteristic such as a NDC for negative bias. We have also observed a shift in gate voltage of the detector's response while varying the QD source-drain voltage from the blockaded region towards the NDC region. We have illustrated that both effects are consistent with variable tunnelling rates in one of the QD's tunnel barriers. The ability to detect modifications in the tunnelling probability of a QD shown here is to be mastered in order to correctly operate these devices for the readout stage in quantum computational schemes.
\\\indent This work was partly supported by Special Coordination Funds for Promoting Science and Technology in Japan. AR acknowledges useful discussions with Dr Gareth Podd. Two of the authors (AR and TF) contributed equally to the work.


%
%





\end{document}